\documentstyle[aps,pra,epsf,twocolumn,floats]{revtex}
\begin{document}
\draft
\twocolumn[\hsize\textwidth\columnwidth\hsize\csname @twocolumnfalse\endcsname

\title{Application of the variational $R$-matrix method to one-dimensional quantum tunneling}
\author{Joseph Kimeu, Roland Mai, and Kingshuk Majumdar}
\address{Department of Physics, Berea College, Berea, KY 40404.}
\date{\today}
\maketitle
\begin{abstract}
{We have applied the variational $R$-matrix method to calculate the reflection and tunneling probabilities of particles tunneling
through one-dimensional potential barriers for five different types of potential profiles -- truncated linear
step, truncated exponential step, truncated parabolic, bell-shaped, and Eckart. 
Our variational results for the transmission and reflection coefficients are compared with exact analytical results and
results obtained from other numerical methods. We find that our results are in good agreement with them. We conclude that the variational 
$R$-matrix method is a simple, non-iterative, and effective method to solve one-dimensional quantum tunneling problems.}
\end{abstract}
\bigskip
]
\narrowtext 

\section {Introduction}
\label{intro}
Quantum tunneling is of considerable interest in modern physics~\cite{raz}. Tunneling of particles through potential
barriers have applications in different areas such as radioactive decay~\cite{ghatak1}, black hole production~\cite{casher},
optical wave-guides, lasers, quantum well structures~\cite{chemla}, and electron transport in micro and nanoscale 
devices~\cite{datta}. In fact spin polarized
tunneling in semiconductor and magnetic heterostructures has opened up a new area in condensed matter 
physics, known as ``spintronics''~\cite{spin}.

The tunneling coefficient is the key element of the tunneling process that must be understood. For example, in the case of nano-devices it 
determines the current-voltage characteristics. In a tunneling problem one usually solves for the tunneling coefficient by solving
the Schr\"{o}dinger equation with appropriate boundary conditions. But 
there are only a few potentials for which the Schr\"{o}dinger equation 
can be solved analytically so that exact results for the tunneling coefficients can be obtained. Other potentials 
must be solved numerically using techniques such as the 
Wentzel-Kramos-Brillouin (WKB)~\cite{ghatak1}, modified conventional WKB (MWKB)~\cite{love}, modified Airy
function (MAF)~\cite{ghatak2}, numerical matrix (MAT)~\cite{matrix}, transfer matrix (TM)~\cite{zhang}, and $R$-matrix~\cite{lane} methods. All these techniques
have been employed to calculate the transmission coefficients for different types of potential barriers.

The $R$-matrix method was originally developed by Wigner for use in nuclear
physics~\cite{wigner}. Later his method was extended by others to solve a variety of different problems in atomic
and molecular physics~\cite{sm,kohn,nesbet,rou}. A variant of the $R$-matrix method known as the variational $R$-matrix 
method, has been successfully applied recently to the problem of quantum tunneling through a potential barrier~\cite{rouzo}.   

In this paper, we report on our test of the effectiveness of the variational $R$-matrix method in obtaining 
the transmission coefficients for a variety of one-dimensional potentials of interest.
We find our variational results to be in good agreement with existing analytical and numerical results.

In Sec.~\ref{form} we briefly present the variational $R$-matrix method. In Sec.~\ref{results} we apply the method for particle 
tunneling through potential barriers. Our results are 
compared to results from analytical expressions and other numerical methods. The results are summarized in Sec.~\ref{end}.        

\section {Formalism} 
\label{form}
In this section, we briefly outline the variational $R$-matrix method following Ref.~\onlinecite{rouzo}. We begin with the 
one-dimensional time independent Schr\"{o}dinger equation for a particle of mass $m$ and with energy $E$ in a potential 
$V(x)$ (assumed to be real),
\begin{equation}
\frac {d^2 \psi}{dx^2} + \frac {2m}{\hbar^2}[E-V(x)]\psi = 0.
\label{sq1}
\end{equation}
As Eq.~\ref{sq1} is a linear and second order differential equation there are two independent solutions, $\psi$ and $\tilde{\psi}$ 
with the same energy $E$.
We define three regions: an inner region ($a\leq x \leq b $) where the potential is a function of the position $x$, and two
outer regions defined by $x\leq a$ and $x \geq b$, where $a$ and $b$ are the boundaries separating  the inner region II from
the outer regions I and III. 

In the regions I and III the potentials $V_1$ and $V_3$ are either zero or constant. So the 
wave-functions ($\psi_1, \tilde{\psi_1}$) and ($\psi_3,\tilde{\psi_3}$) are oscillatory and can be written 
in terms of exponential functions,
\begin{eqnarray}
\psi_1(x) &=& a_1 \exp( ik_1 x) + b_1 \exp (-i k_1 x), \label{psi1}
\\
\tilde {\psi_1}(x) &=& \tilde {a}_1 \exp(ik_1 x) + \tilde{b}_1\exp (-ik_1 x), \label{psi11} \\
\psi_3(x) &=& a_3 \exp( ik_3 x) + b_3 \exp (-i k_3 x), \label{psi3}\\
\tilde {\psi_3}(x) &=& \tilde {a}_3 \exp(ik_3 x) + \tilde{b}_3\exp (-ik_3 x) \label{psi31},
\end{eqnarray}
where $k_1 = \sqrt{2m(E-V_1)}/\hbar$ and $k_3 = \sqrt{2m(E-V_3)}/\hbar$. We assume $E>V_1,V_3$, so $k_1$ and $k_3$ are real.
The coefficients $a_i,\tilde{a}_i,b_i,\tilde{b}_i$ ($i=1,3$) can be
calculated by matching the wave-functions and their first derivatives at the boundaries, i.e. at $x=a$ and $x=b$.

In region II, we expand the wave-function in terms of
a suitable set of $N$  basis functions $\chi_i (x)$, 
\begin{equation}
\psi_2 (x) = \sum_{i=1}^N c_i \chi_i (x). 
\label{psi2}
\end{equation}
Here $c_i$'s are the variational coefficients. In all the cases we have studied we have used
$\chi_i(x) =  \cos(\kappa_i x)$ to be a simple basis set. The parameters $\kappa_i$'s are chosen by trial and error so that the basis
functions $\chi_i(x)$ are sufficiently complete for our input energy values.

We set-up the $R$-matrix for region II by multiplying both sides of Eq.~\ref{sq1} with $\psi_2(x)$ and
then integrate by parts to obtain,
\begin{eqnarray}
&-&\int_a^b \!dx [\psi_2'(x)]^2 + \frac {2m}{\hbar^2}\int_a^b \!dx [E-V(x)]\psi_2^2 (x) \nonumber \\
&+& \lambda (b) \psi_2^2 (b) - \lambda (a) \psi_2^2 (a) = 0.
\label{sq2}
\end{eqnarray} 
In Eq.~\ref{sq2} we have defined the logarithmic derivatives of $\psi_2$ at $x=a,b$: $\lambda (a)= \psi_2'(a)/\psi_2(a)$, 
$\lambda (b)= \psi_2'(b)/\psi_2(b)$. 
We substitute Eq.~\ref{psi2} in Eq.~\ref{sq2}, which now takes the form
\begin{equation}
Q = \sum_{i,j=1}^N \left[A_{ij}c_i c_j + \lambda (b) \Delta_{ij}^b c_i c_j - \lambda (a)\Delta_{ij}^a c_i c_j\right] =0,
\label{Qvar} 
\end{equation}
where ${\bf A}, {\bf \Delta^a}$, and ${\bf \Delta^b}$ are $N \times N$ real symmetric matrices defined as
\begin{eqnarray}
A_{ij} &=& - \int_a^b \!dx \; \chi_i'(x) \chi_j'(x) + \frac {2mE}{\hbar^2}\int_a^b\! dx \; \chi_i (x) \chi_j (x) \nonumber \\
&-& \frac
{2m}{\hbar^2}\int_a^b\! dx \;\chi_i (x) V(x) \chi_j(x), \\
\Delta_{ij}^a &=& \chi_i (a) \chi_j (a), \\
\Delta_{ij}^b &=& \chi_i (b) \chi_j (b).
\end{eqnarray} 
$Q$ in Eq.~\ref{Qvar} does not change by varying $c_i$'s. Thus  by taking the derivative of $Q$ with respect to 
$c_i$ and setting it to zero we obtain a set of equations that can be succinctly written in matrix  form as
\begin{equation}
\left[ {\bf A} + \lambda (b) {\bf \Delta^b} \right] {\bf \cdot C} = \lambda (a) {\bf \Delta^a \cdot C}.
\label{eigen}
\end{equation}
$C$ is a column vector with elements $c_i$'s. 
Eq.~\ref{eigen} is a generalized eigenvalue equation which
should yield exactly $N$ real solutions. But the matrices ${\bf \Delta^a}$ and ${\bf \Delta^b}$ have rank one due to their peculiar 
structures~\cite{rouzo}, so Eq.~\ref{eigen} has a unique solution. We solve Eq.~\ref{eigen}
using the computational software MATLAB to obtain the eigen-vectors ${\bf C}$ and  ${\bf {\tilde C}}$ for two different 
values of the logarithmic derivatives $\lambda (b)$ and ${\tilde \lambda}(b)$. 
Note that the values of the logarithmic derivatives $\lambda(\ell)$ and ${\tilde \lambda}(\ell)$ at a given point 
$x=\ell$ can be chosen arbitrarily~\cite{rouzo}. In this paper, we have chosen $\ell=b$ for our calculations.

Thus, in region II, we obtain two linearly independent
solutions $\psi_2 (x) = \sum_{i=1}^N c_i \chi_i (x)$ and ${\tilde \psi}_2 (x) = \sum_{i=1}^N {\tilde c}_i \chi_i (x)$. 
Next we use the continuity of the wave-functions and their first derivatives at the boundaries
$x=a$ and $x=b$. These yield a set of four simultaneous equations 
\begin{eqnarray}
a_1e^{ik_1 a} + b_1e^{-ik_1a} &=& \sum_{i=1}^N c_i \chi_i (a), \\
ik_1(a_1e^{ik_1 a} - b_1e^{-ik_1a}) &=& \sum_{i=1}^N c_i \chi_i^\prime (a), \\
a_3e^{ik_3 b} + b_3e^{-ik_3 b} &=& \sum_{i=1}^N c_i \chi_i (b), \\
ik_3(a_3e^{ik_3 b} - b_3e^{-ik_3 b}) &=& \sum_{i=1}^N c_i \chi_i^\prime (b),
\end{eqnarray}
which we solve to evaluate the coefficients $a_1,b_1,a_3$, and $b_3$. We repeat the same steps to obtain the other set of
coefficients  ${\tilde a}_1,{\tilde b}_1,{\tilde a}_3$, and ${\tilde b}_3$. These coefficients and $c_i$'s completely
determine the two independent wave-functions $\psi (x) $ and ${\tilde \psi}(x)$ for the entire region. We can now write the general
solution for the entire region as a linear combination of these wave-functions, i.e. 
$\Psi (x) = B \psi (x) + {\tilde B} {\tilde \psi}(x)$ where $B$ and ${\tilde B}$ are two
arbitrary constants.

Next, we obtain expressions for the reflection and transmission coefficients. In
region I, using Eqs.~\ref{psi1} -- \ref{psi31}, the general solution can be written as
\begin{eqnarray}
\Psi_1 (x) &=& B \psi_1 (x) + {\tilde B}{\tilde \psi}_1 (x), \nonumber \\
	   &=& (Ba_1 + {\tilde B}{\tilde a}_1)\left[e^{ik_1 x} + \rho e^{-ik_1 x}\right].
\end{eqnarray}
$\rho$ is the amplitude for reflection and is given by 
$\rho = (Bb_1 + {\tilde B}{\tilde b}_1)/(Ba_1 + {\tilde B}{\tilde a}_1)$. 
Similarly for region III,
\begin{eqnarray}
\Psi_3 (x) &=& B \psi_3 (x) + {\tilde B}{\tilde \psi}_3 (x), \nonumber \\
	   &=& \left[ Ba_3 + {\tilde B}{\tilde a}_3\right]e^{ik_3 x} + 
	   \left[ Bb_3 + {\tilde B}{\tilde b}_3\right]e^{-ik_3 x}.
\end{eqnarray}
We consider the physical situation where we have an incident wave of unit amplitude (which implies 
$Ba_1 + {\tilde B}{\tilde a}_1=1$) coming from the left that partially penetrates the barrier and gets transmitted to region III. 
As there is no reflected wave in region III, we have another constraint $Bb_3 + {\tilde B}{\tilde b}_3 = 0$ on the coefficients $B$ and 
${\tilde B}$. By solving these equations for $B$ and ${\tilde B}$ we obtain the reflection and transmission 
coefficients. They are given by 
\begin{eqnarray}
R &=& |\rho|^2 = \Biggl|\frac {b_1{\tilde b}_3 - b_3{\tilde b}_1}{a_1{\tilde b}_3 - b_3{\tilde a}_1}\Biggr|^2,\\
T &=& \left(\frac {k_3}{k_1}\right)\Biggl|\frac {b_3{\tilde a}_3 - b_3{\tilde a}_3}{a_1{\tilde b}_3 - b_3{\tilde a}_1}\Biggr|^2.
\end{eqnarray}

A good check of the numerical accuracy of our calculation is to calculate the average energy $E_{\rm av}$ in region II ($a \leq x \leq b$)
with the input value of the energy $E$. $E_{\rm av}$ can be evaluated using the expression
\begin{equation}
E_{\rm av} = \frac {\sum_{i,j=1}^N c_ic_jP_{ij}}{\sum_{i,j=1}^N c_ic_j R_{ij}},
\label{aven}
\end{equation}
where 
\begin{eqnarray}
P_{ij} &=& \int_a^b\!dx \left[-\frac {\hbar^2}{2m}\chi_i(x)\chi_j^{\prime \prime}(x) + \chi_i(x) V(x) \chi_j(x)\right], \\
R_{ij}&=& \int_a^b\!dx \chi_i(x)\chi_j(x).
\end{eqnarray}

\section{Applications and Results}
\label{results} 

In order to test the effectiveness of the variational $R$-matrix method we have chosen five 
different one-dimensional potentials: 
truncated linear step, truncated exponential step, truncated parabolic, bell-shaped, and Eckart. We consider particles 
coming from region I with total energy $E$ ($E>V_1,V_3$) striking the potential barrier. According to quantum mechanics, there is a
finite probability for the particles to be transmitted to region III even if the particles have energy less than the maximum
height of the potential barrier. Particles that are not transmitted to region III are reflected back to region I. 
We use the variational method to calculate the
reflection and transmission coefficients as a function of particle energies for each potential barrier. 

For all cases we have checked the accuracy of our calculations by comparing 
the average energies calculated from Eq.~\ref{aven} with the actual input energies. We also have compared our results with exact 
analytical results (ANA) and also with other numerical methods, such as numerical matrix  (MAT)~\cite{matrix}, 
transfer matrix (TM)~\cite{zhang}, Wentzel-Kramos-Brillouin
(WKB)~\cite{ghatak1,ghatak2}, modified conventional WKB (MWKB)~\cite{love}, modified Airy functions (MAF)\cite{ghatak1}, and 
improved MAF (MMAF)~\cite{roy}.

\subsection{Truncated linear step potential}
\begin{figure}[httb]
\protect \centerline{\epsfxsize=2.5in \epsfbox {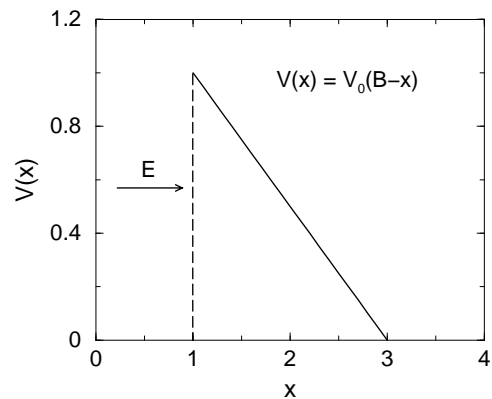}} \vskip .3cm 
\protect \caption{Truncated linear step potential with $V_0=0.5$ hartree, $B=3$ bohr, and boundaries at $x=1$ bohr and $x=3$ bohr.} 
\protect \label{linearfig}
\end{figure}

The truncated linear step potential barrier is a practical example of a potential 
encountered in planar wave-guides~\cite{sodha}. It is defined by the potential function
\begin{equation}
V(x) =\left\{
\begin{array}{l}
0, \quad \quad \quad \quad \quad x<a \\
V_0(B-x), \quad a\leq x \leq b \\
0, \quad \quad \quad \quad \quad x>b
\end{array}\right.
\end{equation}
which is shown in Fig.~\ref{linearfig}. 

For the truncated linear step potential the Schr\"{o}dinger equation can be solved analytically to obtain the 
wave-function~\cite{landau}. 
The exact solution for the inner region can be written in terms of the Airy functions $Ai(x)$ and $Bi(x)$~\cite{as},
\begin{eqnarray}
\psi_2 (x) &=& a_2 Ai\left(\alpha^{1/3}\left[B-x-\epsilon\right]\right) \nonumber \\ 
&+& b_2 Bi\left(\alpha^{1/3}\left[B-x-\epsilon\right]\right),
\end{eqnarray}
where $\alpha=2mV_0/\hbar^2, \epsilon = E/V_0$, and $a_2$ and $b_2$ are coefficients to be determined from the boundary conditions. One can then obtain the exact expressions for the 
reflection ($R_e$) and tunneling ($T_e$) coefficients (see Appendix \ref{app}). 

\begin{figure}[httb]
\protect \centerline{\epsfxsize=3in \epsfbox {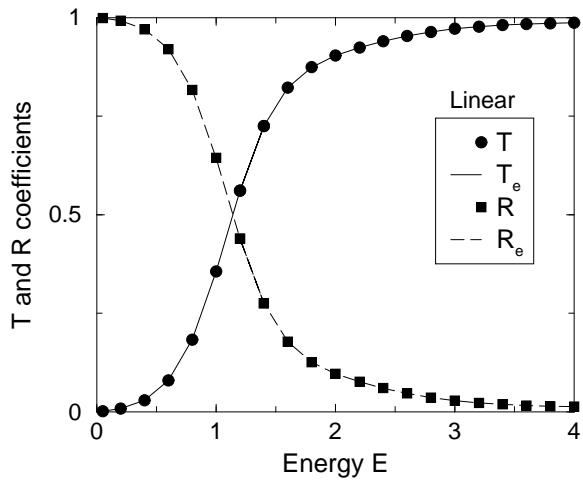}} \vskip .3cm 
\protect \caption{Transmission $T$ (filled circles) and reflection $R$ (filled squares) coefficients through the truncated 
linear step potential 
barrier as a function of particle energy $E$. The solid (long-dashed) lines are for the exact values of $T_e$ ($R_e$) 
obtained from the analytical expression. For the wave-functions $\psi_2(x)$ and ${\tilde \psi}_2(x)$ we have used $\lambda(b)=1$,
${\tilde \lambda}(b)=4$, and $\kappa_i=\{0.1:0.1:6.0\}$(bohr)$^{-1}$ as our input parameters.} 
\protect \label{linear}
\end{figure}

The reflection and tunneling coefficients calculated from our variational $R$-matrix
and analytical methods are plotted in Fig.~\ref{linear} for different values of particle energy $E$ (in units of hartree). 
As $E$ increases it becomes easier for the particles to
tunnel through the barrier -- thus the tunneling coefficient increases (and
the reflection coefficient decreases as $R+T = 1$). 
For $E\approx 1.1$ hartree, the reflection and tunneling probability curves cross each other. At
this value of energy the probabilities of reflection and transmission are each equal to 50\%. 
When the energy of the particles are just over 
the maximum height, $V_0B$ of the truncated linear step potential, i.e. $E \approx 1.5$ hartree, we would expect classically to 
see 100\% transmission. 
But from Fig.~\ref{linear} we find the transmission  probability to be about 
77\%. This reflection of particles for energies greater than the barrier height is also a 
quantum mechanical effect just as the transmission of particles for energies less than the barrier height is a quantum mechanical 
effect.  

As a check of the accuracy of our calculation we have calculated   
the average energy $E_{\rm av}$ using Eq.~\ref{aven} and compared with the actual input energy $E$ and have found that errors 
are in the range 0.03\% -- 1.75\%. The higher errors occur for smaller values of the input energy, which suggests that more
$\kappa_i$'s are required.

In Table~\ref{tab1} we have compared the tunneling coefficients for different values of $B$ obtained from our variational 
method (VRM) with results from other methods. The reason for choosing different values of $B$ for the truncated linear step 
potential is to compare with existing data from other numerical methods~\cite{zhang,roy}. It is known that the
semi-classical WKB approximation fails close to the turning points  (i.e. whenever $E-V(x) \approx 0$) -- so those points are not
taken into account in the WKB calculation. Hence, changes in the $B$ values have no effect on the transmission coefficient. 
In the modified WKB (MWKB) approximation the turning points are included and thus 
the transmission coefficient changes with changes in the $B$ values. However the results are 5\% -- 14\% higher than the expected
results. The variational results are in excellent agreement with the exact results and results from the numerical 
transfer matrix (TM), matrix (MAT), and modified Airy functions (MAF) methods. A good agreement of MAF
results with the exact results is expected, as for the truncated linear step potential the Airy functions are the exact solutions
of the Schr\"{o}dinger equation.

\begin{table}
\caption{Comparison of the variation of tunneling coefficients for different values of $B$ 
with different methods for truncated linear step potential.} 
\begin{tabular}{|c||c||c|c|c|c|c|c|}\hline
$B$ & VRM & ANA &  TM & MAT & WKB & MWKB & MAF \\ \hline \hline
3 & 0.6079 & 0.6077 & 0.6077 & 0.6075 & 0.4670 & 0.6962 & 0.6078 \\
\hline
6 & 0.4819 & 0.4818 & 0.4818 & 0.4815 & 0.4670 & 0.5163 & 0.4856 \\
\hline
8 & 0.4374 & 0.4370 & 0.4370 & 0.4374 & 0.4670 & 0.4522 & 0.4377 \\
\hline
15 & 0.3424 & 0.3423 & 0.3423 & 0.3416 & 0.4670 & 0.3353 & 0.3431 \\
\hline
20 & 0.3053 & 0.3052 & 0.3052 & 0.3048 & 0.4670 & 0.2917 & 0.3050 \\
\hline
\end{tabular}
\protect \label{tab1}
\end{table}

\subsection{Truncated exponential step potential}
\begin{figure}[httb]
\protect \centerline{\epsfxsize=2.5in \epsfbox {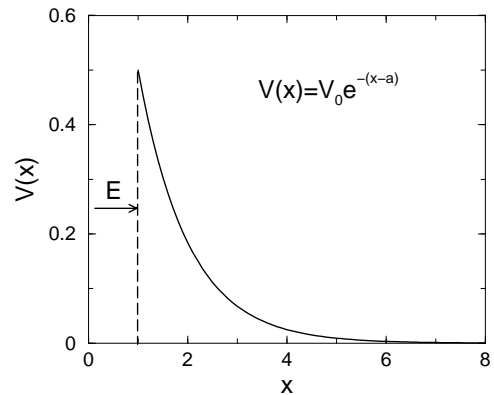}} \vskip .3cm 
\protect \caption{The truncated exponential step potential with $V_0=0.5$ hartree. The potential is truncated at $a=1$ and $b=8$ bohr.}   
\protect \label{expfig}
\end{figure}

The truncated exponential step function, shown in Fig.~\ref{expfig}, is another example  
encountered in planar wave-guides~\cite{sodha}. The potential is characterized by the function
\begin{equation}
V(x) =\left\{
\begin{array}{l}
0, \quad \quad \quad \quad \quad x<a \\
V_0 e^{-(x-a)}, \;\quad a\leq x \leq b \\
0. \quad \quad \quad \quad \quad x>b
\end{array}\right.
\end{equation}
  
For this potential an analytical solution of the Schr\"{o}dinger equation exists, which can be written in terms of the modified
Bessel functions of first and second kinds $I(\nu, z)$ and $K(\nu, z)$~\cite{as},
\begin{eqnarray}
\psi_2 (x) &=& a_2 K\left(2ik,2\alpha^{1/2} e^{-(x-a)/2}\right) \nonumber \\
 &+& b_2 I\left(-2ik,2\alpha^{1/2}e^{-(x-a)/2}\right),
\label{ana_exact} 
\end{eqnarray}
where $k=\sqrt{2mE/\hbar^2}$and $\alpha=2mV_0/\hbar^2$.
Using these wave-functions we have calculated the exact values of $T_e$ and $R_e$ (see Appendix \ref{app}). 
Fig.~\ref{expo} shows the results of our variational calculation and the analytical calculation. 
The reflection and tunneling probability curves cross each other for $E\approx 0.13$ hartree. 
Over the maximum height, $V_0$ of the potential, i.e. $E \approx 0.5$ hartree we find 
from Fig.~\ref{expo}, the transmission  probability to be about 
92\%. Compared to the truncated linear step potential the transmission coefficient increases at a faster rate 
for the truncated exponential step function. This is expected as the slope of the truncated exponential potential decreases as
$-(V_0/a ) \exp (-x/a)$ whereas the slope of the truncated linear step potential remains constant at $-V_0$.  
\begin{figure}[httb]
\protect \centerline{\epsfxsize=3in \epsfbox {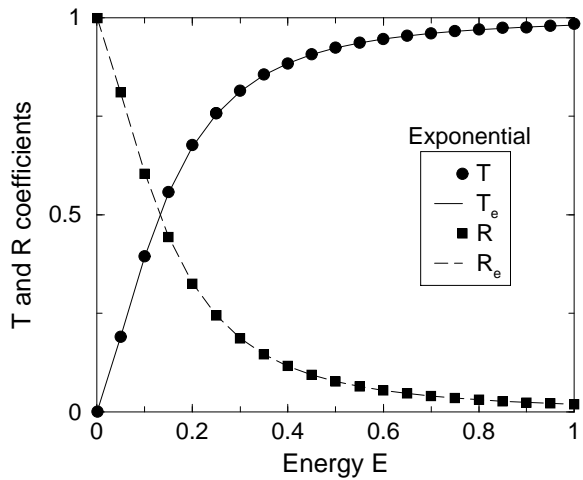}} \vskip .3cm 
\protect \caption{Transmission $T$ (filled circles) and reflection $R$ (filled squares) coefficients through the truncated exponential
step potential 
barrier. Solid (long-dashed) lines are for the exact values of $T_e$ ($R_e$) obtained from the 
analytical
expression. Different input parameters are $\lambda(b)=2$, ${\tilde \lambda}(b)=8,$ and $ \kappa_i=\{0.1:0.1:6.0\}$.} 
\protect \label{expo}
\end{figure}

We have compared the calculated $E_{\rm av}$ values with the input $E$ values to check the accuracy of our calculation and have
found that the error is between 0.013\% -- 2.1\% of the input energy values. This again shows that a denser 
grid of $\kappa_i$'s is needed for smaller $E$ values.

Our variational results (VRM) are compared to other methods in Table
\ref{tab2}~\cite{zhang,roy}. Here we find that the numerical transfer matrix (TM) and matrix (MAT) methods give the same 
results as the analytical results. 
An inspection of Table~\ref{tab2} shows that the results of our variational 
$R$-matrix method are within 0.56\% of the analytical results (ANA); however, one should note that the agreement
is improved as $E$ increases. 
Other methods such as WKB and MWKB again fail to provide satisfactory results. Although MAF results are much more accurate than the
WKB and MWKB results but it is less accurate (differs from exact results by 0.35\% -- 1.5\%) than our variational results.
 
\begin{table} \caption{Comparison of the variation of tunneling coefficients for different values of $E/V_0$ with different methods for 
truncated exponential step potential.}  
\begin{tabular}{|c||c|c|c|c|c|c|c|}\hline
$E/V_0$ & VRM & ANA & TM & MAT & WKB & MWKB & MAF \\ \hline \hline
0.25 &  0.4812 & 0.4789 & 0.4789 & 0.4788 & 0.2247 & 0.3892 & 0.4865 \\
\hline
0.50 & 0.7566 & 0.7549 & 0.7549 & 0.7549 & 0.4221 & 0.8443 & 0.7601 \\
\hline
0.75 & 0.8712 & 0.8702 & 0.8702 & 0.8695 & 0.5693 & 0.9861 & 0.8733 \\
\hline
\end{tabular}
\protect \label{tab2}
\end{table}

\subsection{Truncated parabolic potential}
\begin{figure}[httb]
\protect \centerline{\epsfxsize=2.5in \epsfbox {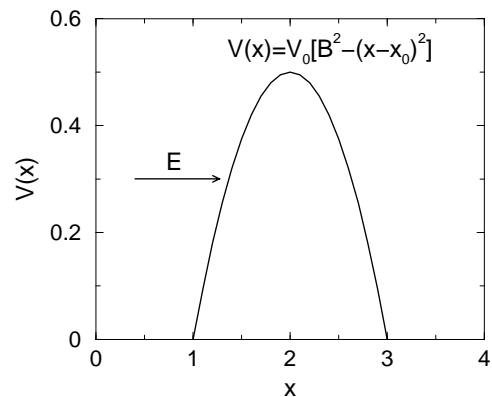}} \vskip .3cm 
\protect \caption{Parabolic potential truncated at $x=x_0 \pm B$ bohr. We have chosen $V_0=0.5$ hartree, $x_0=2$ bohr and $B=1$ bohr for our
calculation.}   
\protect \label{parafig}
\end{figure}

The truncated parabolic potential illustrated in Fig.~\ref{parafig} is of considerable interest in fiber and integrated
optical waveguides~\cite{sodha}. The profile of the potential is given by,
\begin{equation}
V(x) =\left\{
\begin{array}{l}
0, \quad \quad \quad \quad \quad \qquad \qquad \; x<a \\
V_0\left[B^2-(x-x_0)^2\right], \quad a\leq x \leq b \\
0. \quad \quad \quad \quad \quad \qquad \qquad \; \; x>b
\end{array}\right.
\end{equation}

\begin{figure}[httb]
\protect \centerline{\epsfxsize=3in \epsfbox {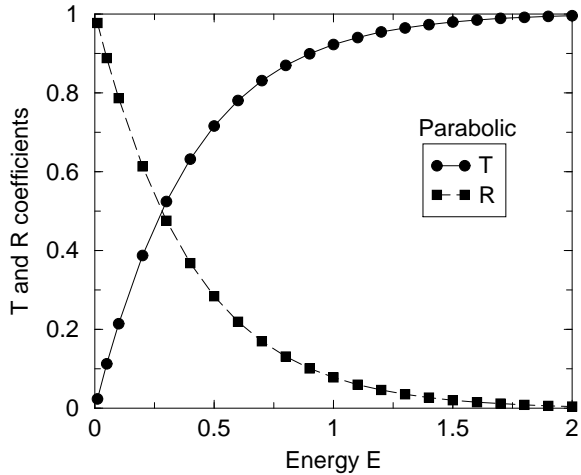}} \vskip .3cm 
\protect \caption{Transmission $T$ (filled circles) and reflection $R$ (filled squares) coefficients through the 
truncated parabolic potential 
barrier.
The exact values for $T_e$ and $R_e$ are not known in this case. The lines are drawn only to guide the eyes.
Input parameters used are $\lambda(b)=2$, ${\tilde \lambda}(b)=3,$ and 
$\kappa_i=\{0.1, 0.5, 0.9, 1.3, 2.0, 2.4, 3.0, 3.4, 4.0, 4.4, 5.0, 5.4, 6.0\}$ (bohr)$^{-1}$.} 
\protect \label{para}
\end{figure}

In Fig.~\ref{para} we have plotted our variational $R$-matrix results for the transmission and reflection coefficients.
Here the particle with $E \approx 0.27$ hartree have equal reflection and transmission probabilities. The transmission
probability is 74\% when the particle energy just exceeds the maximum height $V_0B^2$ of the barrier.   

When compared to our $E_{\rm av}$ values with the input $E$ values we obtained a satisfactory agreement as the errors are 
within 2\%. Our variational results are compared with results from other numerical methods for three different
values of $E/V_0$ in Table \ref{tab3}~\cite{zhang,roy}. The exact analytical results for $T_e$ and $R_e$ are unavailable in this
case. This comparison show that our results from variational $R$-matrix method is in 
excellent agreement with transfer matrix and matrix methods. In this case also, WKB, MAF, and MMAF does not
provide any satisfactory results. 
\begin{table} \caption{Comparison of the variation of tunneling coefficients for different values of $E/V_0$ with different methods for 
truncated parabolic potential.}  
\begin{tabular}{|c||c|c|c|c|c|c|}\hline
$E/V_0$ & VRM & TM & MAT & WKB & MAF & MMAF \\ \hline \hline
0.10 & 0.1122 & 0.1124 & 0.1122 & 0.0575 & 0.0390 & 0.0971 \\
\hline
0.20 & 0.2139 & 0.2141 & 0.2141 & 0.0778 & 0.0506 & 0.1859 \\
\hline
0.50 & 0.4599 & 0.4604 & 0.4603 & 0.1878 & 0.1003 & 0.3981 \\
\hline
\end{tabular}
\protect \label{tab3}
\end{table}

\subsection{Truncated bell-shaped potential}
Our next example is the smooth bell shaped potential defined by  
\begin{equation}
V(x) = \frac {V_0}{\cosh^2 (x-x_0)},
\end{equation}
where $x_0$ is a constant. 
 
\begin{figure}[httb]
\protect \centerline{\epsfxsize=2.5in \epsfbox {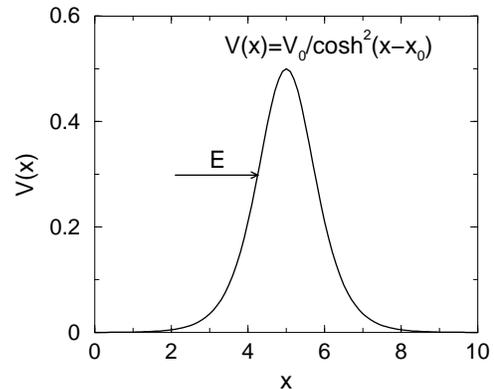}} \vskip .3cm 
\protect \caption{Bell-shaped potential with $V_0=2$ hartree and $x_0=5$ bohr. Inner region is defined between $x=1$ bohr and $x=9$ bohr. 
Potentials $V_1$ in region I and $V_3$ in region III are assumed to be zero.}   
\protect \label{bellfig}
\end{figure}
\begin{figure}[httb]
\protect \centerline{\epsfxsize=3in \epsfbox {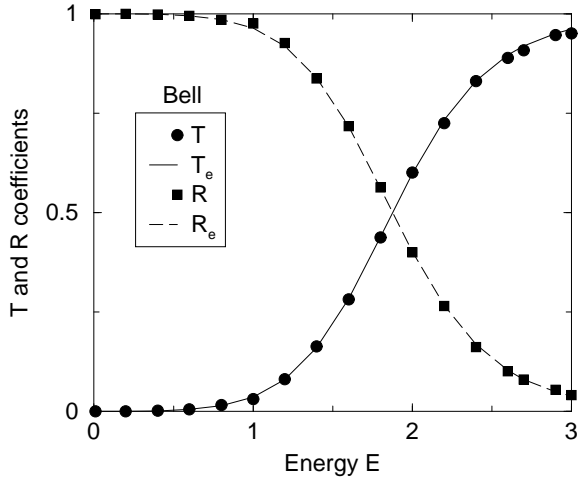}} \vskip .3cm 
\protect \caption{Transmission $T$ (filled circles) and reflection $R$ (filled squares) coefficients through the 
bell-shaped potential 
barrier. Solid and long-dashed lines are for the exact values of $T_e$ and $R_e$. 
Input parameters used are $\lambda(b)=1$, ${\tilde \lambda}(b)=9,$ and $ \kappa_i=\{0.1:0.2:3.0\}$.} 
\protect \label{bell}
\end{figure}

An exact solution of the wave-function of the particle in region II for this potential barrier is known and can be written in terms of the
hypergeometric functions~\cite{landau}. Transmission and reflection coefficients can then be calculated. The exact tunneling
coefficient is given by (for $8V_0> 1$, which is our case) the closed form expression,
\begin{equation}
T_e = \frac{\sinh^2 (\pi k)}{\sinh^2 (\pi k)
+\cosh^2\left(\pi\beta/2 \right)},
\label{bellexact}
\end{equation}
where $k=\sqrt{2mE/\hbar^2}$ and $\beta = \sqrt{8mV_0/\hbar^2-1}$.
In this case our calculated $E_{\rm av}$ values differ from the input $E$ values by 0.0065\% -- 0.694\%.
For the truncated bell-shaped potential we are unable to compare with other numerical methods as we have not found any 
existing numerical results. Fig.~\ref{bell} illustrates our variational results and the exact results. At larger values of the energy $E$ our values of $T$ and $R$
deviate slightly from the exact values which again suggests that  
a better and optimized choice of $\kappa_i$' is necessary to obtain better accuracy. 
From Fig.~\ref{bell} we find that the transmission coefficient is about 62\% when $E$ is over the peak of the barrier. 

\subsection{Eckart's potential}
\begin{figure}[httb]
\protect \centerline{\epsfxsize=2.5in \epsfbox {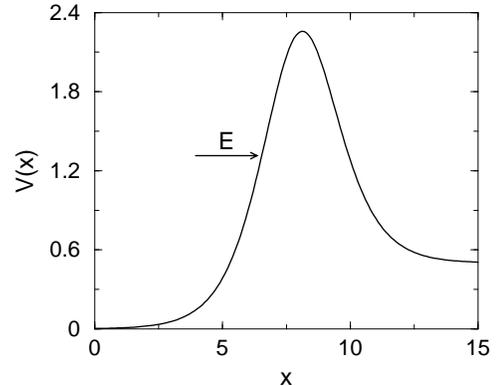}} \vskip .3cm 
\protect \caption{Eckart potential with $A=1$, $B=8$, and $x_0=8$ bohr. 
Inner region is defined between $x=2$ bohr and $x=13$ bohr. 
Potentials $V_1$ and $V_3$ are approximately constants and their values are evaluated at the boundary
values, i.e. $V_1 = V(x)|_{x=1}$ and $V_3 = V(x)|_{x=13}$. The particle energies $E$ are taken to be larger than $V_1$ and $V_3$.}   
\protect \label{eckartfig}
\end{figure}

\begin{figure}[httb]
\protect \centerline{\epsfxsize=3in \epsfbox {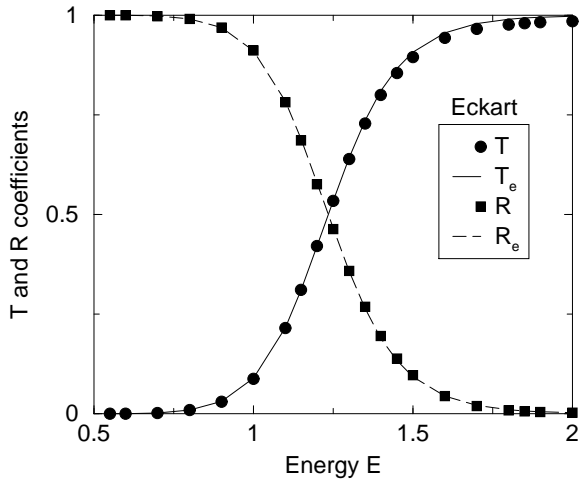}} \vskip .3cm 
\protect \caption{Transmission $T$ (filled circles) and reflection $R$ (filled squares) coefficients for the Eckart potential.
Solid and long-dashed lines  are for the exact values of $T_e$ and $R_e$ obtained from the 
analytical expression in Eq.~\ref{anaeck}. Input parameters used for this calculation are 
$\lambda(b)=2$, ${\tilde \lambda}(b)=13,$ and 
$\kappa_i=\{0.001:0.2:3.0\}$.} 
\protect \label{eckart}
\end{figure}

Eckart introduced a smoothly varying potential function defined as
\begin{equation}
V(x) = \frac {1}{2} \left[ A \frac {e^{x-x_0}}{1+e^{x-x_0}} + B\frac {e^{x-x_0}}{\left(1+e^{x-x_0}\right)^2}\right],
\label{eck}
\end{equation}
where $A$ and $B$ are two dimensionless arbitrary parameters~\cite{eck}. The shape of the potential 
is shown in Fig.~\ref{eckartfig}. 
The Schr\"{o}dinger equation for this potential leads to a differential equation of hypergeometric
type, whose solutions can be written down in terms of the hypergeometric functions from  
which the exact coefficients of reflection and transmission can be extracted~\cite{raz,eck}. The exact expression of $R_e$ is
given by (for $B>1/4$)
\begin{equation}
R_e = \frac {\cosh(2\pi(k-\beta)) + \cosh(2\pi \delta)}
{\cosh(2\pi(k+\beta)) + \cosh(2\pi \delta)},
\label{anaeck}
\end{equation}
where $k=\sqrt{2mE/\hbar^2},\beta = \sqrt{k^2 -A}$, and $\delta = \sqrt{B-1/4}$.
The coefficient $T_e$ is obtained using the relation $T_e = 1 -R_e$. 
The computed values of $T, R$ and the exact values $T_e, R_e$ are shown in Fig.~\ref{eckart}. Our
variational results for $T$ and $R$ are in good agreement with the exact results. From Fig.~\ref{eckart} we find that almost all the
particles are transmitted to region III when their energies are just above the maximum height of the barrier, which is at 2.23 hartree.   
For large values of $E$ there are some discrepancies between the two results which could be improved by enlarging and optimizing the
set of $\kappa_i$. Note that in this case also we have not found any data using other numerical techniques for comparison. For this
case, the errors for $E_{\rm av}$ compared to the input $E$ values are within 0.078\% -- 1.43\%. 

\section{Conclusion}
\label{end}
We have applied the variational $R$-matrix method to calculate the reflection and tunneling probabilities of particles
tunneling through five one-dimensional potential profiles -- truncated linear step, truncated exponential step, truncated parabolic,
bell shaped, and Eckart. We have compared our results with results from analytical and other numerical methods and found
satisfactory agreement with the analytical results and some of the numerical results obtained using numerical matrix and
transfer matrix techniques. Our results can be improved by a better choice of a larger and optimized basis set. We have checked
this by enlarging the basis sets for the truncated linear and exponential potential steps. 
Although semi-classical WKB method is a simple method but it is restricted to slowly varying potentials that are continuous. For potentials
with discontinuities WKB approximation fails. In modified WKB approximation this deficiency is taken into
consideration, but still it fails to provide satisfactory results for any general potential. Modified Airy function and improved 
MAF methods are suitable only for certain potentials such as the truncated linear step, truncated parabolic, and truncated quartic
potentials. Although the matrix method and the transfer matrix technique yield satisfactory results but they are
complicated. On the other hand the variational $R$-matrix method is simple, general, easy to implement, and applies to any 
potential shape.  Also it is non-iterative and thus computationally efficient. 
This method can be applied in scattering theory and also in quantum tunneling in two and three dimensions.

The authors thank S.~T. Powell and A.~Lahamer for helpful discussions and suggestions and gratefully acknowledge the support by
the Undergraduate Research and Creative Projects Program of Berea College.

\section{Appendix}
\label{app}
We assume that for a potential $V(x)$ the  Schr\"{o}dinger equation (Eq.~\ref{sq1}) can be solved exactly for the inner region 
defined by $a \leq x \leq b$ and the solution is given in terms of two functions $f(x)$ and $g(x)$ as
\begin{equation}
\psi_2 (x) = a_2 f(x) + b_2 g(x).
\end{equation}
For example, for the truncated linear step potential the functions $f(x)$ and $g(x)$ are the Airy functions $Ai(x)$ and $Bi(x)$ 
and 
for the truncated exponential step potential they are the modified Bessel functions $K(\nu,z)$ and $I(\nu, z)$.
We now match the wave functions $\psi_1(x), \psi_2(x)$ ($\psi_2(x), \psi_3(x)$) and their derivatives at $x=a$ (at $x=b$) and solve
the set of simultaneous equations to obtain the coefficients $a_i$ and $b_i$'s ($i=1,2,3$). Using these coefficients an exact
expression for the tunneling coefficient can be obtained which is given by, 
\begin{equation}
T_e = 4\left(\frac {k_3}{k_1} \right)\Big|\frac {N}{D}\Big|^2,
\label{exactTe}
\end{equation}
where the numerator $N$ and the denominator $D$ are,
\begin{eqnarray*}
N &=& \left[ f(b) g^\prime (b) - g(b) f^\prime (b) \right],\\
D &=& \Big[g^\prime(b) - ik_3 g(b)\Big]\left[1-\frac {i}{k_1} f^\prime (a) \right] \\
&-& \Big[f^\prime(b) - ik_3 f(b)\Big]\left[1-\frac {i}{k_1} g^\prime (a) \right].
\end{eqnarray*}


\begin{references}
\bibitem{raz} M.~Razavy, {\em Quantum Theory of Tunneling}, World Scientific, NJ, 2003.
\bibitem{ghatak1} A.~K.~Ghatak and S.~Lokanathan, {\em Quantum Mechanics: Theory and Applications}, 3rd. Ed., New Delhi; McMillan, 1984.
\bibitem{casher}A.~Casher and F.~Englert, Class. Quantum Grav. {\bf 10} 2479, (1993).
\bibitem{chemla} D.~S.~Chemla and A.~Pinczuk, Eds. IEEE J. Quantum Electron., {\em Special Issue on Semiconductor Quantum Wells and
Superlattices: Physics and Applications}, {\bf QE-22}, 1986. 
\bibitem{datta} S.~Datta, {\em Electronic Transport in Mesoscopic Systems}, Cambridge Univ. Press, N. Y, 1995.
\bibitem{spin} D.~D.~Awschalom, D.~Loss, and N.~Samarth (Eds.), {\em Semiconductor Spintronics and Quantum Computation}, Springer, N. Y.,
2002.
\bibitem{love} J.~D.~Love and C.~Winkler, J. Opt. Soc. Am. {\bf 67}, 1627 (1977).
\bibitem{ghatak2} A. K. Ghatak, R. L. Gallawa, and I. C. Gayal, IEEE J. Quantum Electron. {\bf 28}, 400 (1992).
\bibitem{matrix} A.~K.~Ghatak, K.~Thyagarajan, and M.~R.~Shenoy, J. Lightwave Technol. {\bf 5}, 660 (1987); 
A.~K.~Ghatak, K.~Thyagarajan, and M.~R.~Shenoy, IEEE J. Quantum Electron. {\bf 24}, 1524 (1988).
\bibitem{zhang} A. Zhang, Z. Cao, Q. Shen, X. Dou, and  Y. Chen, J. Phys. A: Math. Gen. {\bf 33}, 5449 (2000).
\bibitem{lane} A.~M.~Lane and R.~G.~Thomas, Rev. Mod. Phys. {\bf 30}, (1958).
\bibitem{wigner} E.~P.~Wigner and L.~Eisenbud, Phys. Rev. {\bf 72}, 29 (1947).
\bibitem{sm} L.~Smrcka, Microstruct. {\bf 8}, 221 (1990).
\bibitem{kohn}W.~Kohn, Phys. Rev. {\bf 74}, 1763 (1948).
\bibitem{nesbet}R.~K.~Nesbet, {\em Variational Methods in Electron-Atom Scattering Theory}, Plenum, N. Y, 1980.
\bibitem{rou}H.~L.~Rouzo and G.~Raseev, Phys. Rev. A {\bf 29}, 1214 (1984).
\bibitem{rouzo} H.~L.~Rouzo, Am. J. Phys. {\bf 71}, 273 (2003).
\bibitem{roy} S. Roy, A. K. Ghatak, I. C. Goyal, and R. L. Gallawa, IEEE J. Quantum Electron. {\bf 29}, 340 (1993).
\bibitem{sodha} M.~S.~Sodha and A.~K.~Ghatak {\em Inhomogeneous Optical Waveguides}, Plenum, New York, 1977.
\bibitem{landau} L.~D.~Landau and E~.M.~Lifshitz, {\em Quantum Mechanics: Non-relativistic Theory}, Addison-Wesley, MA., 1958.
\bibitem{as} M.~Abramowitz and I.~A.~Stegun {\em Handbook of Mathematical Functions}, Dover Publications, New York, 1970.
\bibitem{eck} C.~Eckart, Phys. Rev. {\bf 35}, 1303 (1930).
\end{references}
\end{document}